\documentclass[sigconf,nonacm]{acmart}
\usepackage{amsmath}
\usepackage{multirow}
\usepackage{xcolor}
\usepackage{subfigure}
\usepackage{appendix}

\begin{document}

\fancypagestyle{firstpagestyle}
{
   \fancyhf{}
   \setlength{\footskip}{7mm}
   \fancyfoot[L]{\footnotesize Workshop for Large Recommender Systems (LargeRecSys), 18th ACM Conference on Recommender Systems, 2024, Bari, Italy.}
}
\thispagestyle{firstpagestyle}

\title{Evaluating Performance and Bias of Negative Sampling in Large-Scale Sequential Recommendation Models}
\author{Arushi Prakash}
\authornote{Both authors contributed equally to this research.}
\affiliation{%
  \institution{Apple}
  \country{}
}
\email{arushi@apple.com}

\author{Dimitrios Bermperidis}
\affiliation{%
  \institution{Apple}
  \country{}
}
\email{d_bermperidis@apple.com}
\authornotemark[1]

\author{Srivas Chennu}
\affiliation{%
  \institution{Apple}
  \country{}
}
\email{srivas.chennu@apple.com}

\renewcommand{\shortauthors}{XXX et al.}

\begin{abstract}
  Large-scale industrial recommendation models predict the most relevant items from catalogs containing millions or billions of options. To train these models efficiently, a small set of irrelevant items (negative samples) is selected from the vast catalog for each relevant item (positive example), helping the model distinguish between relevant and irrelevant items. Choosing the right negative sampling method is a common challenge. We address this by implementing and comparing various negative sampling methods—random, popularity-based, in-batch, mixed, adaptive, and adaptive with mixed variants—on modern sequential recommendation models. Our experiments, including hyperparameter optimization and 20x repeats on three benchmark datasets with varying popularity biases, show how the choice of method and dataset characteristics impact key model performance metrics. We also reveal that average performance metrics often hide imbalances across popularity bands (head, mid, tail). We find that commonly used random negative sampling reinforces popularity bias and performs best for head items. Popularity-based methods (in-batch and global popularity negative sampling) can offer balanced performance at the cost of lower overall model performance results. Our study serves as a practical guide to the trade-offs in selecting a negative sampling method for large-scale sequential recommendation models. Code, datasets, experimental results and hyperparameters are available at: https://github.com/apple/ml-negative-sampling.

\end{abstract}

\begin{CCSXML}
<ccs2012>
   <concept>
       <concept_id>10002951.10003317.10003347.10003350</concept_id>
       <concept_desc>Information systems~Recommender systems</concept_desc>
       <concept_significance>500</concept_significance>
       </concept>
   <concept>
       <concept_id>10010147.10010257.10010258.10010259.10003268</concept_id>
       <concept_desc>Computing methodologies~Ranking</concept_desc>
       <concept_significance>500</concept_significance>
       </concept>
 </ccs2012>
\end{CCSXML}

\ccsdesc[500]{Information systems~Recommender systems}
\ccsdesc[500]{Computing methodologies~Ranking}

\keywords{Recommender Systems, Information Retrieval, Negative Sampling, Sequential Models, SASRec}

% \received{30 Aug 2024}
% \received[revised]{30 Aug 2024}
% \received[accepted]{30 Aug 2024}

\maketitle

\section{Introduction}

Large-scale recommender systems are designed to identify relevant items from vast corpora, such as videos that users might watch or products they may purchase. The underlying models are trained to distinguish between relevant (positive) and irrelevant (negative) items by exposing the model to both during training. Positive items are those with which users have interacted (e.g., liked, purchased, viewed), while negative items are those they have not. Ideally, training would involve all irrelevant items; however, this is computationally infeasible given the scale of the corpora. To mitigate this, negative sampling methods are employed to select a representative subset of negative items. Although this sub-sampling can sometimes degrade modeling performance, as noted by Petrov et al. \cite{Petrov_2023}, it is a necessary trade-off to enable large-scale model training within computational constraints.

\begin{figure*}[hbt]
\includegraphics[width=\textwidth]{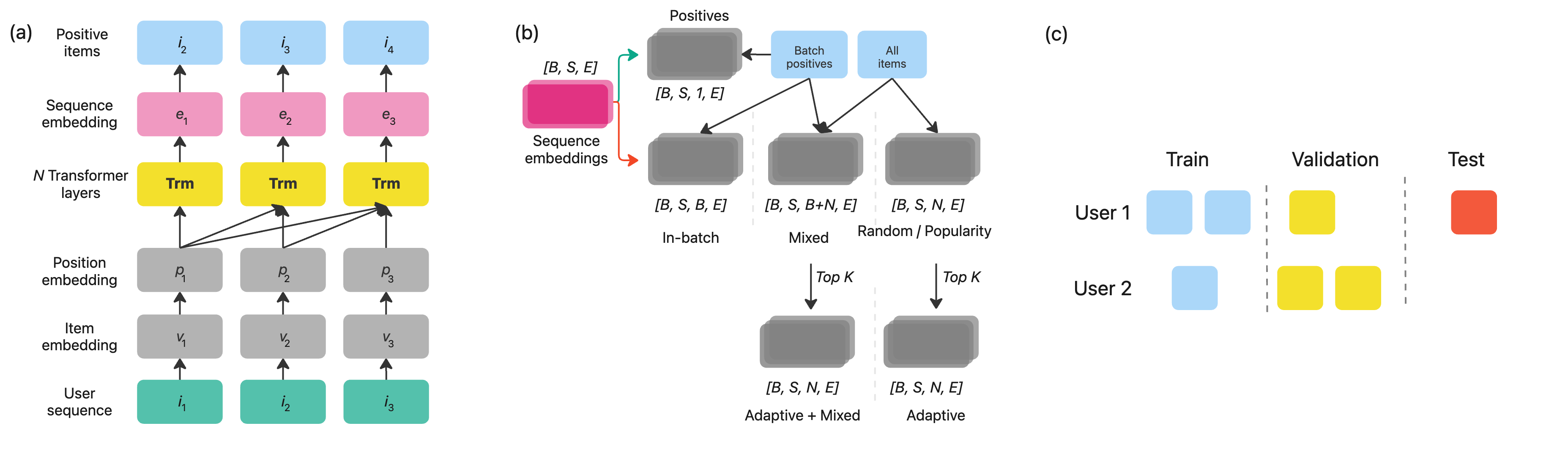}
\caption{(a) Structure of the self-attention sequential recommendation (SASRec) model (b) Structure of positive and negative sample tensors, based on different negative sampling methods, where $B$ is the batch size, $S$ is the sequence length, $N$ is the number of negatives, and $E$ is the dimension of the item embedding (c) Global temporal data splitting applied on benchmark datasets to prevent information leakage}
\label{fig:visual1}
\Description[Structure of the self-attention sequential recommendation (SASRec) model ]{Structure of the self-attention sequential recommendation (SASRec) model}    
\end{figure*}

Negative sampling methods can be categorized into \textit{global} (random \cite{chen2023revisiting}, popularity \cite{mikolov2013distributed}), \textit{local} (in-batch \cite{yi2019sampling, chen2020simple}, cross-batch \cite{wang2021cross}, dynamic \cite{zhang2013optimizing, lai2024adaptive}), and \textit{mixed} (batch-mix \cite{fan2023batch}, mixed \cite{yang2020mixed}) methods. A global method such-as random negative sampling selects negatives uniformly from the entire corpus, while a batch method such in-batch negative sampling samples negatives from the current training batch based on their occurrence probability. In-batch methods are often preferred for their scalability because they do not rely on an external item index to sample negatives.

Negative sampling methods help balance bias and variance in models \cite{yang2020mixed}, targeting issues like popularity and selection bias. Abdollahpouri et al. \cite{abdollahpouri2019unfairness} showed that popularity bias, which is the over-exposure of popular items to users even when they are interested in niche items, can occur when recommendation models are trained on datasets which contain a few, extremely popular items. Methods like popularity-based or batch negative sampling can reduce popularity bias \cite{wu2024effectiveness}. However, sampling based on popularity can lead to selection bias, where less popular items are under-sampled, reducing accuracy for these items \cite{yang2020mixed}. Uniform and mixed sampling methods can alleviate such concerns \cite{yang2020mixed}. Additionally, some methods may retrieve "easy" negatives, which are easily distinguishable from positives and less effective for training. Adaptive sampling methods target "hard" negatives \cite{rendle2014improving}, which are items the model is likely to confuse with positives. Some practitioners use self-supervision \cite{grill2020bootstrap, lee2021bootstrapping} to address these challenges.

These aforementioned studies underscore the importance of carefully selecting negative sampling methods to train effective recommendation models, balancing scalability, bias, and variance. Lyu et al. \cite{lyu2023towards} emphasize considering both dataset statistics and model architecture when choosing a sampling method, presenting results for matrix-factorization and graph-based models. We extend this argument to sequential model architectures, specifically using a self-attentive sequential recommendation (SASRec) model. Sequential models for recommender systems have used either random (SASRec\cite{kang2018self}, TiSASRec\cite{li2020time}, TransRec\cite{wang2022transrec}, FMLPRec\cite{zhou2022filter}, DuoRec\cite{qiu2022contrastive}, Caser\cite{tang2018personalized}, RecJPQ\cite{petrov2024recjpq}), popularity (GRU4Rec+\cite{hidasi2018recurrent}), or in-batch (FEARec\cite{du2023frequency}, ECL-SR\cite{zhou2023equivariant}, GRU4Rec\cite{hidasi2015session}) negative sampling. We also expand the slate of negative sampling methods to six methods.

\section{Contribution}
Our major contributions include:
\begin{itemize}
\item \textbf{Implementing scalable sampling methods for sequential recommendation models.} We have implemented six negative sampling methods namely: random, in-batch, popularity, mixed, adaptive, and adaptive with mixed using the SASRec model as the backbone sequential recommender system model. This enhancement allows the model to handle multiple negative sampling methods and various numbers of negative samples rather than relying on a single random negative sample per positive item. 
\item \textbf{Popularity-aware recommendation metrics.} We demonstrate that averaging model performance metrics over all users can obscure popularity bias. To address this, we have developed recommendation metrics that account for item popularity, providing a more accurate measure of the effectiveness of negative sampling methods across the popularity spectrum.
\item \textbf{Experiments on public datasets.} We conduct extensive offline experiments using publicly available datasets (MovieLens 10M, Amazon Beauty, RetailRocket) to evaluate the effectiveness of different negative sampling approaches. Our work enables practitioners to choose the most appropriate approach based on their dataset statistics and model design.
\end{itemize}

To our knowledge, this is the first comprehensive study of negative sampling methods for sequential models that evaluates their effectiveness in relation to item popularity statistics.

\section{Related work}
\textbf{Global sampling:} Global methods select negative samples based on the overall properties of items, independent of the current training batch or model status. In Random Negative Sampling (RNS) \cite{chen2023revisiting}, negative samples are randomly chosen from the entire pool of items. In contrast, Popularity negative sampling  \cite{mikolov2013distributed} selects negative samples based on item popularity. Although PNS can lead to faster model convergence compared to random sampling, it tends to over-penalize popular items, causing the model to inaccurately recommend less popular items \cite{yang2024does}.

\textbf{Local sampling:} Local methods sample items based on their distribution within the current training batch. In in-batch negative sampling \cite{yi2019sampling, chen2020simple}, negative items are sampled from the current batch. Cross-batch negative sampling \cite{wang2021cross} samples items from both the current and previous batches. Both in-batch and cross-batch sampling methods produce negative samples that approximately follow the global popularity distribution, similar to popularity sampling, and thus face similar issues. Yi et al. \cite{yi2019sampling} recommend using large batch sizes (e.g., 1024 or 8192) and applying sampling correction in the loss function to prevent over-penalizing popular items, similar to popularity sampling. Dynamic \cite{zhang2013optimizing} or adaptive \cite{lai2024adaptive, rendle2014improving} negative sampling methods select negative examples that are ranked close to either the user or the label, representing "hard" examples that the model can learn from effectively.

\textbf{Mixed sampling:} Mixed sampling methods combine global and local sampling strategies. In this approach, Yang et al. \cite{yang2020mixed} combine samples from in-batch and random negative samples in a $1:8$ ratio to create an expanded pool of negative samples, thereby offsetting the popularity bias inherent in purely in-batch items. Chen et al. \cite{chen2022cache} extend in-batch by using a cache of previously informative negative samples. Fan et al. \cite{fan2023batch} propose generating synthetic negatives from batch negatives, with a correction for item popularity to address selection bias.

Given the trade-offs associated with each negative sampling method, Lyu et al. \cite{lyu2023towards} introduced AutoSample, an approach that automatically selects the best sampling method during training, focusing on matrix factorization and graph-based model designs. Our approach complements their work by comparing negative sampling methods with state-of-the-art sequential recommendation models.

\section{Modeling Framework}
In this section, we provide a mathematical formulation of the ranking task followed by the formulation of sampling methods implemented in this paper.

\subsection{Problem Formulation}

The next-item prediction task in recommendation systems aims to predict the next item that the user will interact with from a corpus of items given their recent history of interactions as input. We adopt the sequential recommendation model architecture of SASRec \cite{kang2018self} for computing scores $s(x,y)$. As shown in Fig. \ref{fig:visual1}a, for each user $u$, the model is given the history $\mathcal{H}_u$ of past interactions with items, $\mathcal{H}_u = (i_1^u, i_2^u, . . . , i_{T-1}^u )$ where $t \in [1,2,...,T]$ denote the ordering of the interactions. For each item in the sequence, the model predicts the next item that users will interact with, specifically $\mathcal{H}_u^{\mathrm{shift}} = (i_2^u, i_3^u, . . . , i_T^u )$. This sequence to be predicted is a "shifted" version of the original sequence \cite{kang2018self}. During the forward pass, the model creates a sequence of causal embeddings of the user $$\left(\mathbf{e}^u_2, \mathbf{e}^u_3,\ldots,\mathbf{e}^u_{T-1}\right),$$  where each causal embedding is given as a function of the previous items $$\mathbf{e}_u^t = f(i_1^u, ..., i_{t-1}^u; \boldsymbol{\theta}),$$ where $\boldsymbol{\theta}$ here denotes all trainable model parameters. Finally, the model generates a score for the item at time $t$ in the sequence with the $t$th item in the corpus as
\begin{equation}
s(\mathbf{e}_u^t, \mathbf{e}_i) = \left(\mathbf{e}_u^t\right)^T \cdot \mathbf{e}_i,
\end{equation}
where $\mathbf{e}_i$ is an embedding representation, learned during model training, for the $i$th item in the item corpus. Note that we treat the problem as a collection of binary classification problems \cite{Petrov_2023}, where the probability that the user $u$ will interact with item $i$ at time $t$ is modeled as a \emph{sigmoid} function:

\begin{equation}
P(i|u;t) = \sigma(s(\mathbf{e}_i,\mathbf{e}_u^t)) = \frac{1}{1 + e^{-s(\mathbf{e}_i,\mathbf{e}_u^t)}}.
\end{equation}
State-of-the-art sequential models for next-item prediction, such as SASRec \cite{kang2018self} and CASER\cite{tang2018personalized}, use this problem formulation \cite{petrov2022systematic}. Alternatively, models such as BERT4Rec \cite{sun2019bert4rec} and ALBERT4Rec \cite{petrov2022systematic} model the probability as a softmax function. During training, the model minimizes the binary cross-entropy (BCE) loss function across the entire training sequence, i.e., the loss function for user $u$ is given as  
\begin{equation}
L_u = \sum_{ (i^+,t) \in \mathcal{H}_u^{\mathrm{shift}}} - \left[ log(\sigma(s(\mathbf{e}_{i^+},\mathbf{e}_u^t))) + \sum_{i\not\in \mathcal{H}_u}log(1 - \sigma(s(\mathbf{e}_i,\mathbf{e}_u^t)))\right]
\label{equation:loss}
\end{equation}

Excluding items present in the user's history, the second term is calculated over all items in the corpus, which makes the calculation prohibitively expensive for a million or billion-scale item corpus. To approximate this term, only a subset of items are sampled from the corpus, based on a sampling strategy.
Thus, the second term in \eqref{equation:loss} can be written as 
\begin{equation}
 \sum_{i^- \in \mathcal{N}_u^t}log(1 - \sigma(s(\mathbf{e}_{i^-},\mathbf{e}_u^t))),
\label{equation:neg_samples}
\end{equation}
where the set of negative samples $\mathcal{N}_u^t$  for user $u$ at position $t$ are drawn i.i.d from a negative sampling distribution, which in the most general case is given as,
\begin{equation}
i^- \sim P(\cdot|\mathbf{e}_u^t,i^+),
\label{equation:sample_dist}
\end{equation}
This means that in the most general case negative sampling depends on both the latest user embedding $\mathbf{e}_u^t$, which encapsulates the past user history up to time $t-1$, as well as the latest item that the user interacted with, which also acts as the positive item $i^+$ during training.
In the following section, we will see how different negative sampling strategies can be seen as special cases of Equation \eqref{equation:sample_dist}.

Note that the BCE loss can be directly applied to the sampling scenario because the probability of items are independent of each other \cite{Petrov_2023}. The original SASRec implementation \cite{kang2018self} sampled one negative item per positive item in $Y_u$, chosen with uniform probability over the item corpus (random negative sampling). With this configuration, SASRec outperformed other models in the original publication. Later studies suggested expanding the number of negatives to improve modeling accuracy \cite{Petrov_2023}.

\subsection{Negative sampling methods}
To implement negative sampling with BCE loss, we created tensors containing the scores $s(x,y)$ for positive and negative items for the first and second terms of Equation \ref{equation:loss}, respectively. Across negative sampling methods, the tensor for scores of positive items for each item in the user sequence is of size $[B, S]$, where $B$ is the size of the training batch, and $S$ is the maximum length of user sequence.

The negative items and tensor for negative items changes according to the negative sampling technique, as shown in Fig. \ref{fig:visual1}b.

\begin{itemize}
\item \textbf{Global (random, popularity):} Excluding items already present in the user sequence, negative items are chosen for each user uniformly at random based on item frequency or inverse item frequency. The negative tensor is of shape $[B, S, N]$ where $N$ is the number of negative items. Each item in the user sequence is scored against all the negative items available for the user. For this family of negative samplers, Equation \eqref{equation:sample_dist} reduces to the unconditional distribution

\begin{equation*}
P(i^-|\mathbf{e}_u^t,i^+) = P(i^-) = \phi(i^-)^\gamma,
\end{equation*}

where $\phi(i^-)$ is the popularity of the negative item and the exponent $\gamma\in\{1,0\}$. \\

\item \textbf{Batch or in-batch:} All positive items in the training batch are considered negative items, excluding items that are present in the user's sequence. The negative tensor is of shape $[B, S, B]$.

\item \textbf{Mixed \cite{yang2020mixed}:} In-batch negatives $[B, S, B]$ and random negatives $[B, S, N]$ are combined into a size $[B, S, B+N]$ tensor.

\item \textbf{Adaptive\cite{rendle2014improving}:} Given a tensor of negatives generated by RNS, ANS only retains the top $K$ highest scoring negative items for each item in the user's sequence. Given randomly sampled negatives $[B, S, N]$, the negative tensor retains the shape $[B, S, N]$ but the values of $N - K$ items are reduced to $0$. Furthermore, the sampling distribution takes the form $P(i^-|\mathbf{e}_u^t,i^+) = P(i^-|\mathbf{e}_u^t)$, i.e., it is conditioned only on the user history as encapsulated by the causal user embedding.

\item \textbf{Adaptive with mixed:} Given a tensor of negatives, arising from any of the above sampling methods, this method only retains the top $K$ highest scoring negative items for each item in the user's sequence. For example, given mixed sample negatives $[B, S, B+N]$, the negative tensor retains the shape $[B, S, B+N]$ but the values of $B + N - K$ items are reduced to $0$. Furthermore, in the adaptive case, the sampling distribution takes the form $P(i^-|\mathbf{e}_u^t,i^+) = P(i^-|\mathbf{e}_u^t)$, i.e., it is conditioned only the user history as encapsulated by the causal user embedding.

\end{itemize}

Table \ref{tab:methods_comparison} provides a summary of the above negative sampling methods.

\begin{table}[hbt]
\centering
\caption{Comparison of negative sampling methods across different criteria}
\label{tab:methods_comparison}
\begin{tabular}{@{}lccc@{}} % l for left-aligned text, c for centered columns
\toprule
 & \textbf{In-batch} & \textbf{Global} & \textbf{Model-based} \\ 
\midrule
PopRec      & - & \checkmark & - \\
RNS            & - & \checkmark & -  \\
PNS         & - & \checkmark & -  \\
BNS           & \checkmark & - & -  \\
MNS              & \checkmark & \checkmark & -  \\
ANS           & \checkmark & - & \checkmark\\
AMNS & \checkmark & \checkmark & \checkmark  \\
\bottomrule
\end{tabular}
\end{table}

\begin{figure}[hbt!]
\centering
\includegraphics[width=0.5\textwidth]{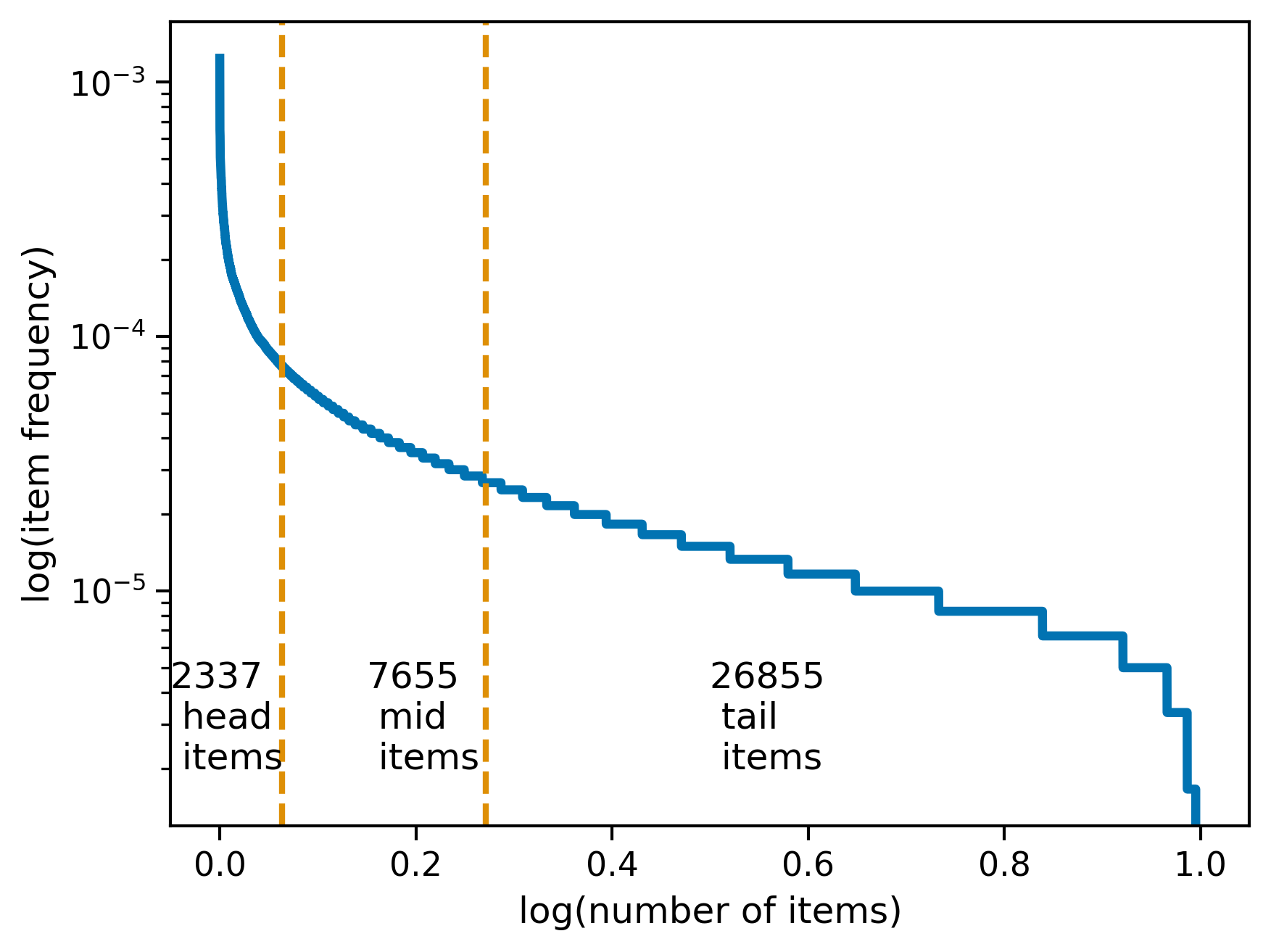}
\caption{Popularity-based cohorts in the RetailRocket dataset}
\label{fig:cohorts}
\Description[Popularity-based cohorts in the Steam dataset]{Popularity-based cohorts in the Steam dataset}
\end{figure}

\subsection{Evaluation}
\label{sec:evaluation}

\textbf{Hit-rate:} We assess the accuracy of all the models produced by different negative sampling methods by adapting hit-rate (HR), originally defined as:
\begin{equation}
\text{Hit Rate @k} = \frac{1}{|\mathcal{U}|}\sum_{u\in\mathcal{U}}\mathbf{1}\{i_u^t\in\mathcal{R}(u;k)\}
\end{equation}

where $\mathcal{R}(u;k)$ are the top-$k$ items retrieved by the recommendation model from the full item corpus. We allow the model to retrieve top-$k$ items from the entire item corpus instead of retrieving top-$k$ from $100$ items from the corpus sampled randomly \cite{kang2018self} or by popularity \cite{dallmann2021case}. This removes inconsistencies between full and sampled metrics \cite{dallmann2021case, krichene2020sampled} and prevents the bias of metrics towards some negative sampling strategies.

We re-formulate the hit-rate (HR) to account for popularity, as:
\begin{equation}
\mathrm{\text{Hit Rate}}_{\text{cohort}}\text{ @k} = \frac{1}{|\mathcal{C}|}\sum_{u\in\mathcal{C}}\mathbf{1}\{i_u^t\in\mathcal{R}(u;k)\}
\end{equation}
where $C$ is the popularity-cohort of the item. An item belongs to one of three cohorts - $C_{head}$, $C_{mid}$, and $C_{tail}$ - based on its frequency of occurrence in the training dataset (see Fig. \ref{fig:cohorts} for an example), resulting in an item distribution shown in Table \ref{tab:datasets}. For the test set, a data point is assigned to head, mid, or tail cohorts based on the last item in the sequence, which serves as the label for the test set.

The above reformulation of HR aligns with the concepts of average coverage of long-tail items \cite{abdollahpouri2019managing} and popularity-stratified recall \cite{steck2011item}, both of which highlight the impact of popularity bias on recommender systems.

\textbf{Balance:} We assess the distribution of accuracy between head, mid, and tail cohorts of all the models produced by different negative sampling methods using a new metric called "balance" and defined as:

\begin{equation}
    \mathrm{Balance} = 1.0 - \mathrm{Gini}([HR_{head}, HR_{mid}, HR_{tail}]),
\end{equation}

where $\mathrm{Gini}(\cdot)$ is the Gini coefficient that measures the dispersion or inequality of a distribution.

\begin{table}[ht]
\centering
\caption{Number of items (\%) and median percentage of users (MPU) who interact with an item, for each popularity cohort and dataset.}
\label{tab:datasets}
\begin{tabular}{@{}llccc@{}}
\toprule
\multirow{2}{*}{\textbf{Cohort}} & & \multicolumn{3}{c}{\textbf{Dataset}} \\ \cmidrule(lr){3-5} 
 &  & \textbf{ML-10M} & \textbf{Beauty} & \textbf{RetailRocket} \\ 
\midrule
\multirow{2}{*}{\textbf{Head}} & \textbf{Items (\%)} & 2.2 & 4.6 & 6.3 \\
 & \textbf{MPU (\%)} & 15.9  & 0.041 & 0.032 \\
\midrule
\multirow{2}{*}{\textbf{Mid}} & \textbf{Items (\%)} & 6.9 & 19.1 & 20.8 \\
 & \textbf{MPU (\%)}  &  5.4 & 0.012 & 0.011  \\
\midrule
\multirow{2}{*}{\textbf{Tail}} & \textbf{Items (\%)} & 90.8 & 76.3 & 72.9 \\
 & \textbf{MPU (\%)}  & 0.13  & 0.003 & 0.003  \\
\bottomrule
\end{tabular}
\end{table}

\begin{figure}[htbp]
\centering
\includegraphics[width=0.5\textwidth]{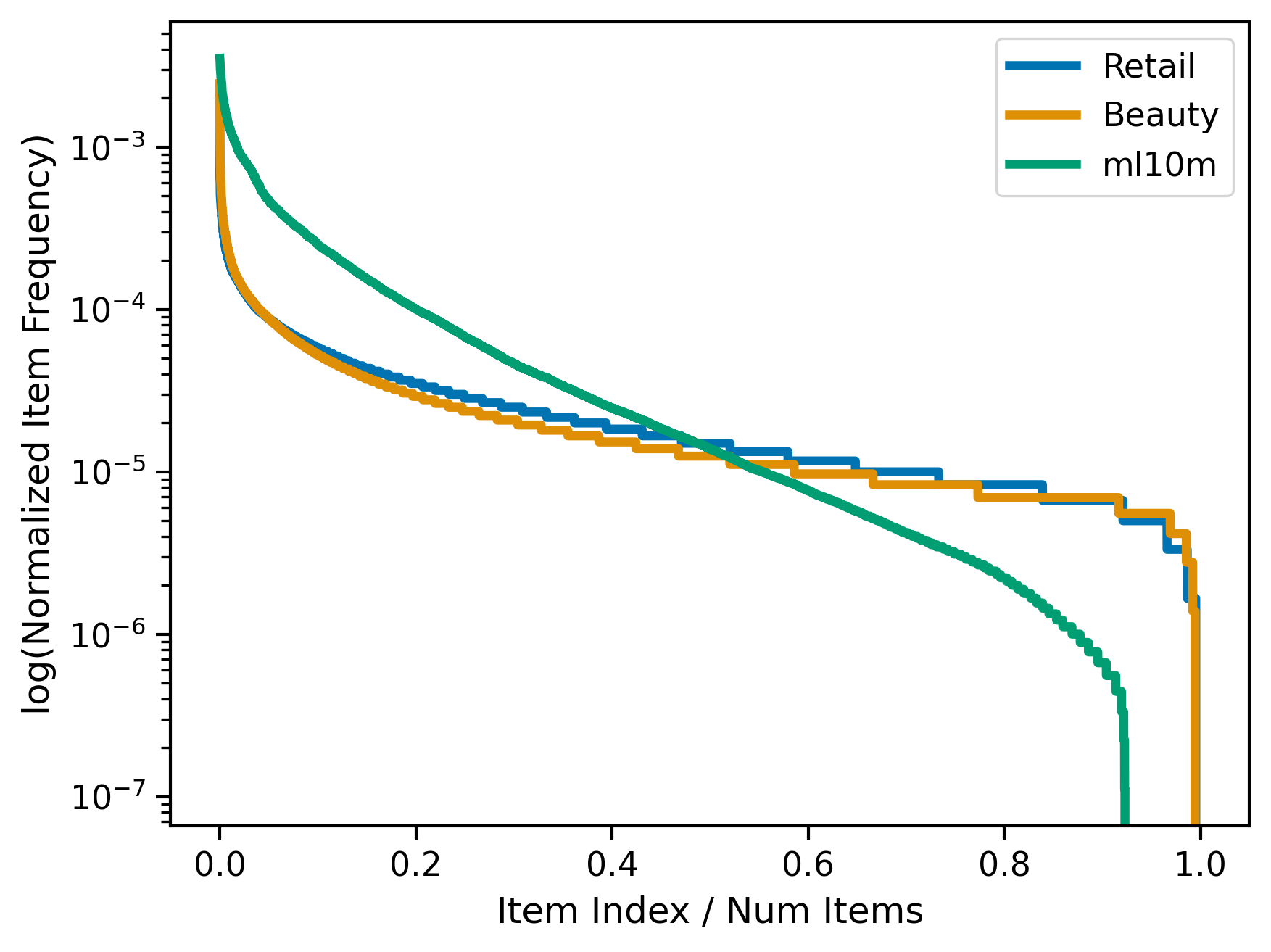}
\caption{Histogram of normalized popularity distributions of the public benchmark datasets, MovieLens 10M, Amazon Video and RetailRocket.}
\Description{Histogram of normalized popularity distributions of the public benchmark datasets, MovieLen 10M, Amazon Video and RetailRocket.}
\label{fig:all_dist}
\end{figure}

\begin{figure*}[htbp]
\centering
\includegraphics[width=\textwidth]{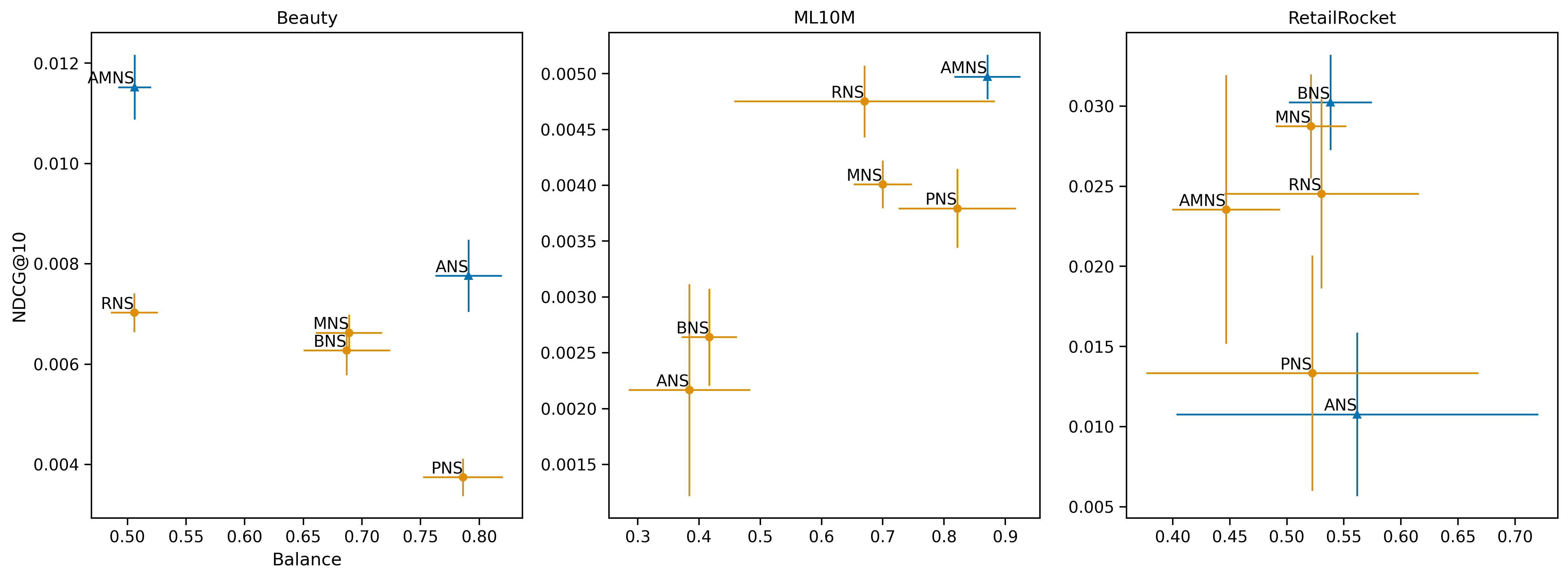}
\caption{Average NDCG@10 for all datasets (left) MovieLens 10M, (center) Amazon Beauty, and (right) RetailRocket for 20x runs for each point}
\label{fig:results}
\Description[Average NDCG@10 for all datasets (left) MovieLens 10M, (center) Amazon Beauty, and (right) RetailRocket for 20x runs for each point]{Average NDCG@10 for all datasets (left) MovieLens 10M, (center) Amazon Beauty, and (right) RetailRocket for 20x runs for each point} 
\end{figure*}

\begin{table*}[h]
\centering
\begin{tabular}{ll|cccc|ccccc}
\hline
\multirow{2}{*}{\textbf{Dataset}} & \multirow{2}{*}{\textbf{Method}} & \multicolumn{4}{c}{\textbf{HR@10}} &  \multicolumn{4}{c}{\textbf{NDCG@10}} \\ \cmidrule(lr){3-6} \cmidrule(lr){7-11}
 &  & \textbf{Total} & \textbf{Head} & \textbf{Mid} & \textbf{Tail} & \textbf{Total} & \textbf{Head} & \textbf{Mid} & \textbf{Tail} & \textbf{Balance}\\ 
\hline
             & PopRec  & 0.91 & 4.00 & 0.00 & 0.00 & 0.44 & 1.91 & 0.00 & 0.00 & 33.3 \\
             & Adaptive w/ mixed    & \textbf{2.33} & \textbf{7.14} & 2.15& 0.27& \textbf{1.15} & \textbf{3.65} & 0.99& 0.12& 50.62 \\
             & Adaptive     & 1.64& 2.15& \textbf{2.63} & \textbf{0.89}& 0.78& 1.01& \textbf{1.26} & \textbf{0.42} & \textbf{79.09} \\
Amazon Beauty & Batch    & 1.37& 2.8& 2.11& 0.34& 0.63& 1.29& 0.97& 0.15& 68.7 \\
             & Mixed     & 1.47& 2.96& 2.36& 0.34& 0.66& 1.33& 1.08& 0.15& 68.89 \\
             & Popularity     & 0.77& 0.9& 1.28& 0.46& 0.37& 0.39& 0.64& 0.23& 78.61 \\
             & Random     & 1.44& 4.41& 1.49& 0.1& 0.7& 2.21& 0.68& 0.04& 50.59 \\
\hline
             & PopRec  & 0.64 & \textbf{5.26} & 0.00 & 0.00 & 0.26 & \textbf{2.13} & 0.00 & 0.00 & 33.3 \\
             & Adaptive w/ mixed    & \textbf{1.1} & 0.8& \textbf{1.05} & \textbf{1.19}& \textbf{0.5}& 0.35& \textbf{0.46} & \textbf{0.54}& \textbf{87.05} \\
             & Adaptive     & 0.46& 0.14& 0.02& 0.66& 0.22& 0.07& 0.01& 0.31& 38.41 \\
ML-10M       & Batch     & 0.58& 0.02& 0.1& 0.84& 0.26& 0.01& 0.04& 0.39& 41.66 \\
             & Mixed     & 0.89& 0.33& 0.48& 1.13& 0.4& 0.14& 0.2& 0.52& 69.96 \\
             & Popularity     & 0.83& 0.54& 0.59& 0.96& 0.38& 0.25& 0.26& 0.44& 82.14 \\
             & Random     & 1.04& 3.06& 0.64& 0.72& 0.48& 1.41& 0.27& 0.33& 67.0 \\
\hline
             & PopRec  & 1.58 & 7.29 & 0.00 & 0.00 & 0.52 & 2.39 & 0.00 & 0.00 & 33.3 \\
             & Adaptive w/ mixed    & \textbf{4.19} & \textbf{15.62} & 2.34& 0.29& 2.36& 8.76& 1.42& 0.12& 44.67 \\
             & Adaptive     & 1.91& 2.34& 3.23& 0.99& 1.08& 1.03& 2.36& 0.39& \textbf{56.16} \\
RetailRocket & Batch     & 4.09& 11.82& \textbf{3.47} & \textbf{1.1}& \textbf{3.02} & \textbf{9.16} & 2.75& \textbf{0.53} & 53.81 \\
             & Mixed     & 3.77& 11.61& \textbf{3.47} & 0.56& 2.87& 8.98& \textbf{2.84}& 0.26& 52.11 \\
             & Popularity     & 2.02& 4.17& 3.19& 0.45& 1.34& 2.44& 2.53& 0.19& 52.21 \\
             & Random     & 3.58& 11.35& 3.14& 0.47& 2.45& 7.7& 2.41& 0.22& 53.04 \\
\hline
\end{tabular}
\caption{Performance metrics of different negative sampling methods with SASRec sequential model backbone across Amazon Beauty, MovieLens 10M, and RetailRocket datasets. These metrics are averaged over 20 runs.}
\label{tab:performance_metrics}
\end{table*}

\section{Experiments}

\subsection{Setup}
\label{sec:Setup}

\quad\textbf{Data:} We performed experiments on three publicly available benchmark datasets -- MovieLens 10M (ML-10M), Amazon Beauty, and RetailRocket \cite{kang2018self, hidasi2023widespread}. These datasets, particularly RetailRocket, exhibit strong sequential patterns, making them highly relevant to a sequential model architecture and evaluation framework.

\textbf{Train-test split:} Recent studies have shown that the leave-one-out methods conventionally used to create training, validation, and test data splits suffer from information leakage \cite{hidasi2023widespread, campos2011towards, meng2020exploring}, leading to inaccurate offline metrics. To prevent this information leakage, we employ a global temporal splitting strategy \cite{meng2020exploring}, as illustrated in Fig. \ref{fig:visual1}c, to generate temporally non-overlapping splits for training, validation, and test data. This approach ensures that the validation and test datasets resemble the scenarios encountered by recommender systems at inference time in real-world systems, where only data up to a certain point in time is available to the model.

\textbf{Optimizing runtime sampling:} Following the setup in \cite{kang2018self}, we generate training mini-batches with negative samples throughout the training process. These batches are placed in a queue to be consumed by one training epoch. For advanced sampling methods (mixed, adaptive with mixed), this generation process can be a bottleneck for model training. To mitigate this, we replace the queue with a fixed size buffer of mini-batches that starts filling up as soon as training begins. After the specified buffer size is reached, the training epoch starts reusing previously stored mini-batches with random probability. We fix the buffer size by specifying an \emph{oversampling factor} (OSF), which is the number of training epochs after which mini-batches are re-used. OSF determines how many times on average a user - positive item pair might participate in the negative sampling process. The OSF hyperparameter introduces an accuracy vs. speed trade-off, where a larger OSF provides higher accuracy but reduces the speed of training. Empirically, we observed that increasing OSF beyond a certain point produces diminishing returns (see Appendix A for an illustration of this effect). Therefore, for each negative sampling method, we selected a value of OSF that would ensure that model training would take a reasonable time to complete while minimally impacting the final accuracy.

\textbf{Hyperparameters:} We retained some of the original model hyperparameters (number of heads, hidden units, embedding size, maximum sequence length, dropout rate) for each dataset. We explored other hyperparameters for the model (batch size, learning rate) and negative sampling (number of negatives, number of mixed negatives, number of adaptive negatives). For every result, we report the test accuracy for the epoch where validation accuracy is maximized. We run each configuration 20 times and report the average and standard deviation of metrics. 

\textbf{Popularity baseline model (PopRec):} We create a baseline model named PopRec, which recommends items based on their popularity. For each dataset, we identify the top $k$ items with the highest frequency in the training data. These items are then recommended to all users. Validation and testing data are excluded to prevent information leakage.

\subsection{Results}

We present the results of each sampling method with each dataset in Table \ref{tab:performance_metrics}, and plot the results in Fig. \ref{fig:results} to illustrate the dispersion in model performance metrics.

\textbf{Model performance depends highly on dataset distribution and negative sampling technique:} In Fig. \ref{fig:all_dist}, we examine the popularity distributions of each dataset using a semi-log plot of the item-frequency versus item index. The frequencies on the $y$-axis are normalized by the total number of interactions to facilitate comparison across datasets. The item frequencies in the Amazon Beauty and RetailRocket datasets exhibit similar distributions characterized by a heavy mid ($~20\%$ of items) and tail ($~74\%$ of items) sections. The ML-10M has a truncated head and mid sections ($<10\%$ of items). Table \ref{tab:datasets} reports the Median Percentage of Users (MPU) who interact with an item within each popularity cohort. The number of users interacting with a head vs tail item is $122:1$ for ML-10M, $14:1$ for Amazon Beauty, and $11:1$ for RetailRocket datasets. This shows that ML-10M dataset contains fewer items with more number of users interacting with each head. The skew is lesser in Amazon Beauty and RetailRocket datasets.

In Fig. \ref{fig:results}, we show the performance of all methods on these datasets. For the ML-10M dataset, which features a high volume of items in the tail (see Table \ref{tab:datasets}) we see that random and adaptive with mixed methods, where most negative samples are drawn using random sampling, are able to achieve both high model performance and balance. Batch sampling performs badly likely due to the over-penalization of frequently occurring popular items and decreased randomness in sampling compared to popularity sampling. For Amazon Beauty and RetailRocket datasets, sampling methods like adaptive with mixed and random sampling are unable to bring "balance" in model performance across popularity cohorts. Further, in the Amazon Beauty dataset, we see that adaptive and popularity negative sampling, which rely on sampling from in-batch popular or global popular negative samples, provide the highest balance but low model performance. We hypothesize that due to the higher volume of items in mid and tail there is a greater volume of users in these datasets who prefer niche items rather than popular items. We refrain from making strong conclusions from the RetailRocket dataset because of the large standard deviations in metric values but present the data to show the reader the difference in performance of sampling methods across datasets.

\textbf{Random negative sampling reinforces popularity bias:} A majority of sequential models in literature (SASRec\cite{kang2018self}, TiSASRec\cite{li2020time}, TransRec\cite{wang2022transrec}, FMLPRec\cite{zhou2022filter}, DuoRec\cite{qiu2022contrastive}, Caser\cite{tang2018personalized}, RecJPQ\cite{petrov2024recjpq}) use randomly selected negative samples which might be perceived as an unbiased method of selecting negatives while maintaining high model performance. We find (Table \ref{tab:performance_metrics} and Fig. \ref{fig:results}) that RNS achieves high model performance but reinforces popularity bias in the data.  The reinforcement leads to the model performing much better for head items than either mid or tail items. Specifically, the ratio of head NDCG to tail NDCG is greater than 1 and greater than in other methods ($44.1$ for Amazon Beauty, $4.25$ for ML-10M, and $24.1$ for RetailRocket). This also aligns with the findings from Chen et al.\cite{chen2023fairly} who showed that random sampling increased the probability of sampling from unpopular items and inherited popularity bias from the data. \\

\textbf{Global popularity sampling penalizes popular items:}  An alternative to random sampling is to sample negative items based on their global popularity\cite{hidasi2018recurrent}. This method can help reduce performance imbalance by increasing HR$@10$ for mid and head cohorts but at the cost of low HR$@10$ for the head cohort. Consequently, it significantly reduces overall model performance compared to random sampling. The decease in NDCG compared to random is $46.5\%$ for Amazon Beauty, $20.9\%$ for ML-10M, and $43.5\%$ for RetailRocket. Notably, popularity sampling shows high "balance" values but consistently underperforms for accuracy (HR$@10$) and ranking (NDCG$@10$) metrics.

\textbf{Adaptive with mixed sampling achieves high model performance:} In Fig. \ref{fig:results}, we see that adaptive with mixed sampling achieves the highest performance across all datasets. Chen et al. also showed that model performance for sequential recommenders improve when negative samples are selected in an adaptive manner based on model's current state and user history instead of choosing negatives via RNS\cite{chen2022generating}. This high performance might be explained by the model's ability to choose "hard" negatives or negatives that the model might easily confuse for positive items. Adaptive with mixed sampling also reinforces popularity bias in the dataset like random sampling owing to most negative coming from random samples the ratio of batch to random samples is $1:10$.

\begin{figure}[htbp]
\centering
\includegraphics[width=0.5\textwidth]{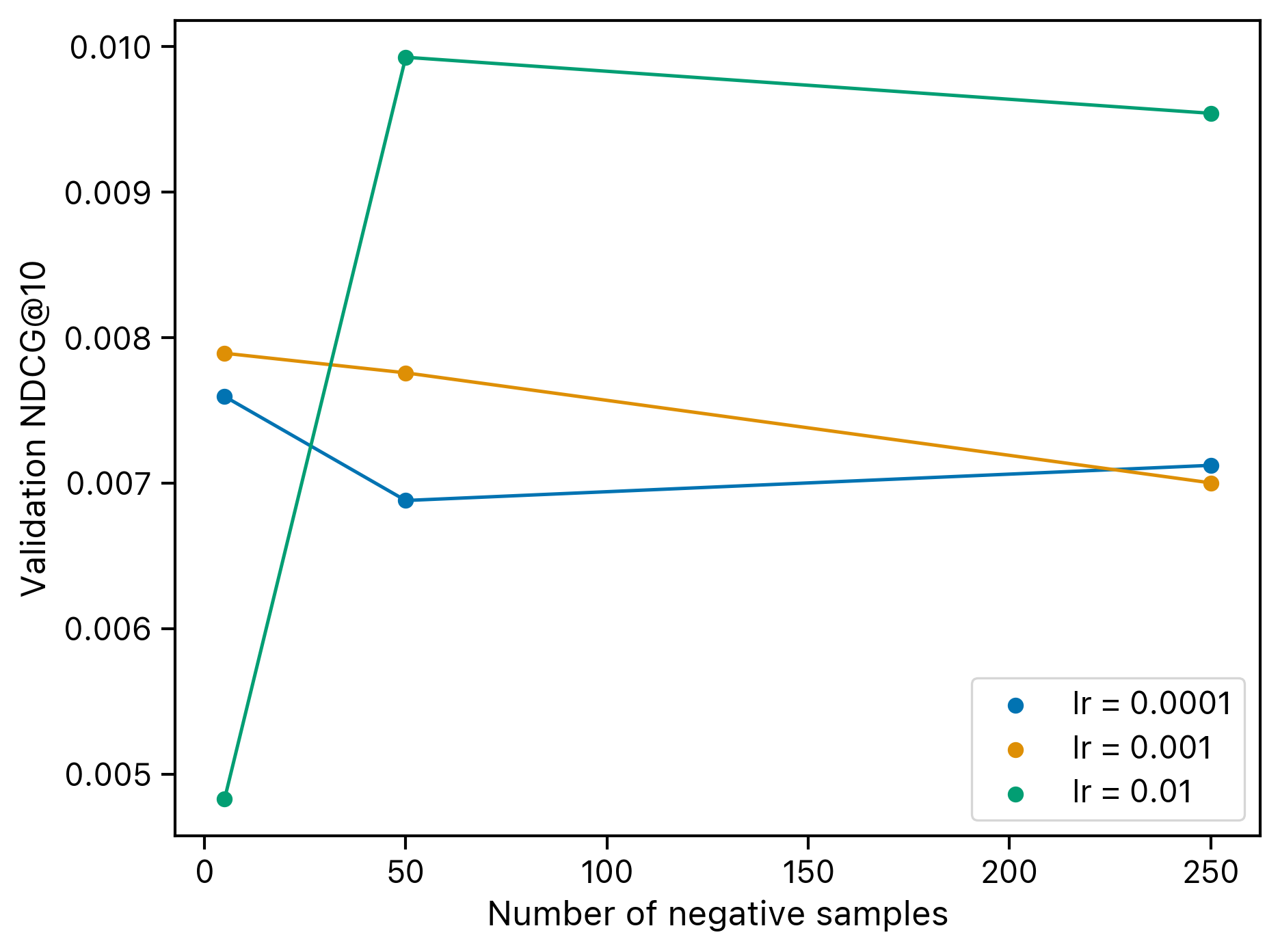}
\caption{NDCG@10 on validation data for the ML-10M dataset as a function of learning rate.}
\label{fig:n_samples}
\Description[NDCG@10 on validation data for the ML-10M dataset as a function of learning rate.]{NDCG@10 on validation data for the ML-10M dataset as a function of learning rate.}
\end{figure}

\textbf{Model performance is a non-linear function of number of negative samples:} We found that model performance varied with hyperparameters, see Fig. \ref{fig:n_samples}. This contradicts Petrov et al. \cite{Petrov_2023} who showed that increasing the number of randomly samples negatives increases the performance of SASRec. This is likely because we used global temporal splitting instead of leave-one-out splitting used by their study, which has been shown to cause potentially incorrect estimation of model performance\cite{meng2020exploring}. Consequently, we urge practitioners to explore all hyperparameters before settling on the appropriate number of negative samples.

\section{Conclusion and future work}

In this paper, we extend established negative sampling methods to the SASRec sequential recommendation models. We examine the performance of these sampling methods through the critical lens of popularity bias, using expanded evaluation metrics to assess the effectiveness of recommendation models on head, mid, and tail items. Our findings are based on three publicly available datasets, which exhibit different popularity distributions. We explore the strengths and limitations of various negative sampling approaches and demonstrate that conventional methods often overemphasize popular items, adversely affecting the diversity of large-scale recommendation systems. Using rigorous and repeatable experiments, we show that novel adaptive with mixed negative sampling can mitigate popularity bias while maintaining overall model performance for a dataset like ML-10M which features extremely heavy interaction with popular items. The impact of our study is to provide valuable insights for practitioners aiming to build large-scale and high-quality recommendation systems in the industry. Future research may focus on extending these methods to other recommendation models and integrating negative sampling into a mixture-of-experts model to optimize sampling at the user level.

\begin{acks}
We would like to thank Matthew Jockers, Sofia Nikolakaki, Riyaaz Shaik and Aditya Nair for their input on drafts of this manuscript.
\end{acks}

\bibliographystyle{ACM-Reference-Format}
\bibliography{references}

\appendix
\appendixpage % Optional title page for appendices
\setcounter{figure}{0}                       % <---------------
\renewcommand\thefigure{A\arabic{figure}}   % <---------------

\section{Sampling Optimization}
\begin{figure}[h]
\centering
\includegraphics[width=0.5\textwidth]{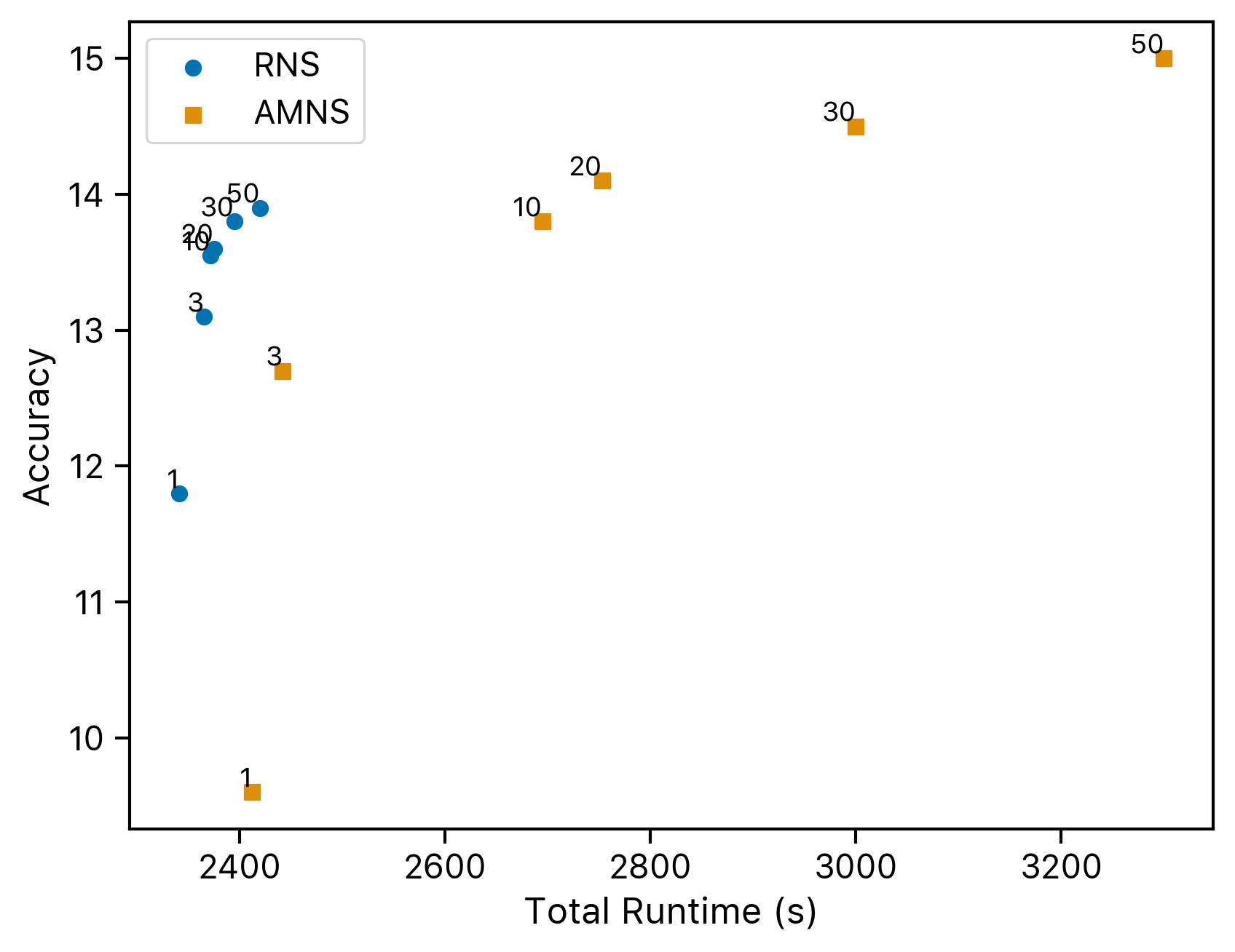}
\caption{Scatter plot comparing random and adaptive with mixed in terms of total accuracy vs runtime trade-off for various sizes of the mini-batch cache (see point labels).}
\Description[Scatter plot comparing random and adaptive with mixed in terms of total accuracy vs runtime trade-off for various sizes of the mini-batch cache]{Scatter plot comparing RNS and AMNS in terms of total accuracy vs runtime trade-off for various sizes of the mini-batch cache}
\label{fig:osf}
\end{figure}

As explained in section \ref{sec:Setup}, we oversample mini-batches of negatives to improve computational efficiency. Figure \ref{fig:osf} depicts the results of experimenting with various buffer sizes for the generated mini-batches. The numbers depicted in the plot show the cache size in terms of oversampling factor (OSF). From the results, we can draw two conclusions:

\begin{enumerate}
    \item First, increasing OSF generally provides diminishing returns in terms of accuracy, while increases the computational load and runtime required.
    \item Second, AMNS is able to better utilize the increased sample diversity provided by larger OSF in order to achieve higher accuracy (at cost of runtime) compared to RNS which reaches a plateau sooner.
\end{enumerate}

\end{document}